\documentclass[prd,showpacs,preprintnumbers]{revtex4} % Physical Review D

\usepackage{graphicx}% Include figure files
\usepackage{dcolumn}% Align table columns on decimal point
\usepackage{bm}% bold math
\usepackage{latexsym}
\usepackage{amsfonts}
\usepackage{amssymb}
\usepackage{amsmath}
\begin{document}

%\preprint{KUNS-}

\title{
Higher Curvature Corrections to Primordial Fluctuations in Slow-roll Inflation
}

\author{Masaki Satoh}
\email{satoh@tap.scphys.kyoto-u.ac.jp}
\author{Jiro Soda}
\email{jiro@tap.scphys.kyoto-u.ac.jp}
\affiliation{
 $^{\ast,\dagger}$Department of Physics,  Kyoto University, Kyoto
 606-8501, Japan
}

\date{\today}% It is always \today, today,

%===============================================================%
%************************* ABSTRACT ****************************%
%===============================================================%
\begin{abstract}
We study higher curvature corrections to the scalar spectral index, 
the tensor spectral index, the tensor-to-scalar ratio, and the polarization
of gravitational waves. 
We find that the higher curvature corrections can not be negligible in the dynamics
of the scalar field, although they are energetically negligible.
Indeed, it turns out that the tensor-to-scalar ratio could be enhanced
and the tensor spectral index could be blue due to the
 Gauss-Bonnet term.
We estimate the degree of circular polarization  of gravitational
waves generated during the  slow-roll inflation. 
We argue that the circular polarization can be observable 
 with the help both of the Gauss-Bonnet and parity violating terms.
 We also present several examples to reveal 
 observational implications of higher curvature corrections 
 for chaotic inflationary models. 
\end{abstract}

%\pacs{98.80.Cq}% PACS, the Physics and Astronomy
\maketitle
\section{Introduction}

According to the Wilkinson Microwave Anisotropy Probe (WMAP)
 data~\cite{Komatsu08} and other observations, 
it seems  incontestable that the large scale structure of the universe
stems from quantum fluctuations during the slow-roll inflation.
 Considering the accuracy of future observations, 
 we expect higher order corrections to the simplest
inflation model become important.
Hence, it would be worth to study the corrections
 due to higher curvature terms to the slow-roll inflation.
To the leading order,  
 the corrections come from the Gauss-Bonnet and the axion-like parity
 violating coupling terms~\cite{Weinberg08}. 
In fact, the same type of corrections can be derived from a particular 
 superstring theory~\cite{Antoniadis:1993jc}. 
 At present, however, we have not fully understood such corrections 
 based on superstring theory.
Hence, we have to resort to the phenomenological approach.
 We hope the understanding of higher curvature corrections
 to the inflationary scenario
 provides a clue to the study of superstring theory.

There have been many works concerning the Gauss-Bonnet and the 
parity violating corrections to the Einstein gravity.
It is  known that Gauss-Bonnet term can provide non-singular 
cosmological solutions~\cite{Antoniadis:1993jc, Kawai:1998bn}
 and also provide alternative explanation 
 for dark-energy~\cite{Nojiri:2005vv,neupane06a,Neupane08,neupane06b}.
There are many works on scalar and tensor perturbations
 in Gauss-Bonnet cosmology~\cite{Kawai:1999xn, Kawai:1999pw, 
 Hwang:1999gf, Cartier:2001is, Kawai:1998ab, Soda:1998tr, 
Kawai:1997mf,Gasperini:1997up,Cartier:2001gc,Leith:2007bu,Guo07,
Vazquez:2008wb}. 
The parity violating term has been considered in conjunction with the  circular
polarization of gravitational
 waves~\cite{Lue:1998mq, Choi:1999zy, Alexander:2004wk, Lyth:2005jf, Satoh08}. 
However, since no one has systematically investigated  both terms simultaneously
in the context of the slow-roll inflation,
 the combined effect on the slow-roll inflation is not so clear.
 Moreover, observational implications have not been discussed so far.
 Hence, a systematic study of higher curvature corrections
 to slow-roll inflationary scenario is desired.
 
 In this paper,  we study the slow-roll inflation with higher curvature corrections
 and derive the formula  for the observables in terms of slow-roll parameters.
 It is stressed that the corrections could be relevant to observables in spite of
  the fact that 
 its energetic contribution can be negligible. The reason is as follows.
 The Gauss-Bonnet term behaves as an effective  potential term in the scalar
 field equation, though it does not contribute to the Friedmann equation 
 in the slow-roll approximation. It reminds us of a hybrid model.
 That is why  the effect could alter the tensor-to-scalar ratio. 
There is also an effect of the
parity violating term on the gravitational waves as the leading effect.
 In fact, the circular polarization will be produced
 during the slow-roll inflation, which never occurs in the conventional
 inflationary scenario. 

The organization of the paper is as follows:
In section II, we define slow-roll parameters in the inflationary scenario
with higher curvature corrections. 
In section III, we calculate the scalar spectral index. 
In section IV, we obtain the tensor spectral index
and the tensor-to-scalar ratio.
Here, we show the possibility of the enhancement of the tensor-to-scalar
ratio due to the Gauss-Bonnet term. 
 We also estimate the degree of the circular polarization
of the primordial gravitational waves. 
In section V, the concrete examples are presented as an illustration.
 The final section is devoted to the conclusion.

\section{Slow-roll inflation}

In this section, we will identify the slow-roll parameters. 
Since the parity violating term is not relevant to the background
evolution, we have the corrections solely from the Gauss-Bonnet term.
What we shall do is to characterize the correction in terms of the 
newly defined slow-roll parameters. 

We consider the gravitational action with
the Gauss-Bonnet and the parity violating terms
\begin{align}
 S&= \frac{M_{\rm Pl}^2}{2}\int{\rm d}^4x\sqrt{-g}R
 -\int{\rm d}^4x\sqrt{-g}\left[
 \frac{1}{2}\nabla^\mu\phi\nabla_\mu\phi
 +V(\phi  )\right] \nonumber\\
 & \qquad \qquad 
 -\frac{1}{16}\int{\rm d}^4x\sqrt{-g} \xi (\phi)R^2_{\rm GB}
 +\frac{1}{16}
 \int{\rm d}^4x\sqrt{-g} \omega (\phi)R\tilde{R} \ ,
 \label{act}
\end{align}
where $M_{\rm Pl}^2\equiv 1/8\pi G$ denotes the reduced Planck mass and
$g_{\mu\nu}$ is the metric tensor. 
The Gauss-Bonnet term $R_{\rm GB}$ and 
the parity violating term $R\tilde{R}$
are defined by
\begin{align}
 &R^2_{\rm GB}\equiv R^{\alpha\beta\gamma\delta}R_{\alpha\beta\gamma\delta}
 -4R^{\alpha\beta}R_{\alpha\beta}+R^2 \ ,
 \\
 &R\tilde{R}\equiv\frac{1}{2}\epsilon^{\alpha\beta\gamma\delta}R_{\alpha\beta\rho\sigma}
 {R_{\gamma\delta}}^{\rho\sigma}  \ .
\end{align}
The inflaton field $\phi$ has the potential $V(\phi)$.
We have introduced coupling functions $\xi(\phi)$ and $\omega(\phi)$.
In principle, these functions should be calculated from
the fundamental theory, such as superstring theory. 
Hence, it might be important to put constraints on these
functions through observations.

Let us take  the flat Friedmann-Robertson-Walker(FRW) metric
\begin{align}
  {\rm d}s^2  =
 -{\rm d}t^2+
 a^2 (t)\delta_{ij}{\rm d}x^i{\rm d}x^j
 \ ,
 \label{frw}
\end{align}
where $a$ is the scale factor.
From the action (\ref{act}), we obtain following background equations
\begin{align}
 & 3M_{\rm Pl}^2H^2 = \frac{1}{2}\dot{\phi}^2 +V
 +\frac{3}{2} {H}^3 \dot{\xi} \label{eq1} \\
 & \ddot{\phi} + 3 {H} \dot{\phi} 
 + \frac{3}{2} {H}^2(\dot{H}+H^2) \xi_{,\phi}
 +V_{,\phi}=0 
 \ ,
 \label{eq2}
\end{align}
where a dot denotes the derivative with respect to the cosmic time, and
 $H$ is the Hubble parameter defined by $H\equiv \dot{a}/a$.
Note that coupling function $\xi$ works as the effective potential 
for the inflaton $\phi$, and only the Gauss-Bonnet coupling $\xi$ appears
in background equations because of parity symmetry of FRW metric.
We will see the effect of the parity violating term appears only 
in tensor perturbations.

We  consider the friction-dominated slow-roll inflation phase
where the following inequalities hold:
\begin{eqnarray}
  \frac{\dot{\phi}^2}{M_{Pl}^2 H^2} \ll 1 \ , \quad
  \frac{H^3 \dot{\xi}}{M_{Pl}^2 H^2} = \frac{H\dot{\xi}}{M_{Pl}^2 } \ll
  1 \ , \quad
  \frac{\dot{H}}{H^2} \ll 1 \ , \quad
  \frac{\ddot{\phi}}{H \dot{\phi}} \ll 1 \ .
 \label{eq11}
\end{eqnarray}
Under these circumstances, Eq.(\ref{eq1}) reads
\begin{align}
 H^2 = \frac{V}{3M_{\rm Pl}^2}\ .
\end{align}
This is nothing but the conventional slow-roll Friedman equation.
While, the scalar field equation (\ref{eq2}) becomes
\begin{align}
  \dot{\phi}
 = -\frac{V_{,\phi}}{3H}
 -\frac{1}{2} H^3 \xi_{,\phi}
 \label{eq22}
 \ .
\end{align}
Here, the higher curvature correction plays a roll. 
If the last term can be neglected, we have the conventional 
slow-roll inflationary scenario. On the other hand, if the last term
becomes important, the situation is different.
Namely, the inflation is driven by the potential $V$, whereas the dynamics
of the scalar field is determined by the Gauss-Bonnet term.

Now, we need to check the consistency conditions.
The consistency condition for (\ref{eq11}) reads
\begin{eqnarray}
 \frac{\dot{\phi}^2}{M_{Pl}^2 H^2} 
 = \frac{M_{Pl}^2 V_{,\phi}^2}{V^2} + \frac{V_{,\phi}\xi_{,\phi}}{36 M_{Pl}^6}
    + \frac{V_{,\phi}\xi_{,\phi}}{3M_{Pl}^2 } \ll 1
\end{eqnarray}
and
\begin{eqnarray}
  \frac{H^3 \dot{\xi}}{M_{Pl}^2 H^2} = \frac{H \xi_\phi \dot{\phi}}{M_{Pl}^2}
  = - \frac{\xi_{,\phi} V_{,\phi}}{3M_{Pl}^2}
    - \frac{H^4 \xi_{,\phi}^2}{2M_{Pl}^2}  \ll 1 \ .
\end{eqnarray}
Third consistency condition for Eq.(\ref{eq11}) leads to
\begin{align}
 \frac{\dot{H}}{H^2}
 =\frac{1}{2H}\frac{(H^2\dot{)}}{H^2}
 =\frac{\dot{\phi}}{2H}\frac{V_{,\phi}}{V} 
 = -\frac{M_{Pl}^2}{2}\left( \frac{V_{,\phi}}{V} \right)^2
    - \frac{1}{12}\frac{ \xi_{,\phi} V_{,\phi}}{M_{Pl}^2}  \ll 1 \ .
 \label{hdot}
\end{align}
Taking time derivative of Eq.(\ref{eq22}), we get another consistency condition
\begin{align}
 \frac{\ddot{\phi}}{H\dot{\phi}}
 &=
 \frac{M_{\rm Pl}^2}{2}\frac{V_{,\phi}^2}{V^2}-
 M_{\rm Pl}^2\frac{V_{,\phi\phi}}{V}
 -\frac{V_{,\phi}\xi_{,\phi}}{4M_{\rm Pl}^2}
 -\frac{V\xi_{,\phi\phi}}{6M_{\rm Pl}^2}  \ll 1
 \ .
\end{align}
For the slow-roll assumption to be valid, these terms must be sufficiently
small.  Hence, 
we  define {\it five} slow-roll parameters,
\begin{align}
 \epsilon\equiv\frac{M_{\rm Pl}^2}{2}\frac{V_{,\phi}^2}{V^2}\ ,
 \quad
 \eta\equiv M_{\rm Pl}^2\frac{V_{,\phi\phi}}{V}\ ,
 \quad
 \alpha\equiv\frac{V_{,\phi}\xi_{,\phi}}{4M_{\rm Pl}^2}\ ,
 \quad
 \beta\equiv\frac{V\xi_{,\phi\phi}}{6M_{\rm Pl}^2}\ ,
 \quad
 \gamma\equiv \frac{V^2\xi_{,\phi}^2}{18M_{\rm Pl}^6}
 =\frac{4}{9}\frac{\alpha^2}{\epsilon}\ ,
\end{align}
and impose the conditions 
$\epsilon,|\eta|,|\alpha|,|\beta|,\gamma\ll 1$. 
If these five conditions are satisfied, 
the inflaton field $\phi$  rolls down the potential slowly.

Note that, due to the Gauss-Bonnet term, the number of parameters and
 accompanying conditions increased from two for the conventional slow-roll
inflation to five for this case.
What we want to know is the effect of new parameters $\alpha, \beta, \gamma$
on the cosmological fluctuations. 

For later convenience, we define $\rho^2$ and $\sigma$ as
\begin{align}
 \rho^2=
 \frac{1}{M_{\rm Pl}^2}
 \frac{\dot{\phi}^2}{H^2}=
 2\epsilon+\frac{4}{3}\alpha+\frac{\gamma}{2}
 ,\quad
 \sigma
 =\frac{H\dot{\xi}}{M_{\rm Pl}^2}=
 -\frac{4}{3}\alpha-\gamma \ .
 \label{rho}
\end{align}
It is also useful to notice the following relation
\begin{align}
 \frac{\dot{H}}{H^2}=
 -\frac{M_{\rm Pl}^2}{2}\frac{V_{,\phi}^2}{V^2}
 -\frac{V_{,\phi}\xi_{,\phi}}{12M_{\rm Pl}^2}
 =-\epsilon-\frac{\alpha}{3}\label{hdot2}  \ .
\end{align}
From the next section, we study corrections to cosmological perturbations
due to higher curvature terms.

\section{Scalar Perturbations}

In this section, we study the higher curvature corrections 
to the scalar perturbations.
It is convenient to use conformal time $\tau$, defined as 
$a(t){\rm d}\tau={\rm d}t$, rather than the cosmic time $t$.

We consider the scalar perturbations $A,B,\psi$, and $E$
defined by the perturbed metric
\begin{align}
  ds^2  = a^2 (\tau) \left[ - (1+2A)d\tau^2 + 
 B_{|i}{\rm d}\eta{\rm d}x^i
 +(\delta_{ij}+2\psi\delta_{ij}+2E_{|ij}){\rm d}x^i{\rm d}x^j \right]
 \ ,
 \label{frw_s}
\end{align}
where the bar denotes the spatial derivative and the prime represents
the derivative with respect to the conformal time. 
Here, we choose the gauge,  $E=\delta\phi=0$.
In this gauge, modified Einstein equations for scalar quantities become
\begin{align}
 &2(1-\sigma/2)\nabla^2\psi-2{\cal H}(1-3\sigma/4)\nabla^2B
 -6{\cal H}(1-3\sigma/4)\psi'=2a^2 VA/M_{\rm Pl}^2-3\sigma{\cal H}^2A
 \\
 &(1-3\sigma/4){\cal H}A=-(1-\sigma/2)\psi'
 \\
 &(1-\sigma/2)A-\left(1-\frac{\ddot{\xi}}{2M_{\rm Pl}^2}\right)\psi
 +\frac{1}{a^2}[(1-\sigma/2) a^2B]'=0\ ,
\end{align}
where $\cal H$ is conformal Hubble parameter, defined by
${\cal H}\equiv a'/a=aH$. 
By eliminating $A$ and $B$ from the above equations,
we can get the single equation 
for $\psi$~\cite{Kawai:1999pw,Hwang:1999gf}.
It is also straightforward to obtain the action for $\psi$
from the result. 

Using Fourier expansion of $\psi$
\begin{align}
 \psi =
 \frac{1}{M_{\rm Pl}}\int\frac{{\rm d}k^3}{(2\pi)^3}
 \psi_{\bf k}{\rm e}^{i{\bf k\cdot x}}\ , 
\end{align}
we can write down the action for $\psi_{\bf k}$
\begin{eqnarray}
 S=\frac{1}{2}\int{\rm d}\tau\int \frac{{\rm d}k^3}{(2\pi)^3}
 A_\psi^2\left(
 |\psi'_{\bf k}|^2
 -C^2_\psi k^2|\psi_{\bf k}|^2
 \right)
  \ ,
\end{eqnarray}
where we have defined 
\begin{align}
 A_\psi^2&\equiv
 6a^2(1-\sigma/2)\left[
 1-\left(1-\rho^2/6-\sigma\right)
 \frac{1-\sigma/2}{(1-3\sigma/4)^2}
 \right] \ , 
 \\
 C_\psi^2&\equiv
 \frac{2}{A_\psi^2}\left\{
 \left[\frac{a(1-\sigma/2)^2}{H(1-3\sigma/4)}\right]'
 -a^2\left(1-\frac{\ddot{\xi}}{2M_{\rm Pl}^2}\right)
 \right\} \ .
\end{align}
Note that, due to the parity symmetry in scalar modes, there never 
appears the effect of the coupling function $\omega(\phi)$ in the action.

 Defining a new variable $\Psi_{\bf k}\equiv A_\psi\psi_{\bf k}$,
we obtain the canonical action
\begin{eqnarray}
 S =\frac{1}{2}\int{\rm d}\tau\int\frac{{\rm d}k^3}{(2\pi)^3}
 \left(
 |\Psi'_{\bf k}|^2
 -C^2_\psi k^2|\Psi_{\bf k}|^2
 +\frac{A''_{\psi}}{A_{\psi}}|\Psi_{\bf k}|^2
 \right)  \ .
\end{eqnarray}
This gives the Schrodinger type equation 
\begin{eqnarray}
 \Psi''_{\bf k}+\left(C_\psi^2k^2-\frac{A_\psi''}{A_\psi}\right)\Psi_{\bf k}=0\ ,
 \label{eom_Psi}
\end{eqnarray}
Notice that the term $A_\psi''/A_\psi$ behaves as an effective potential, and
$C_\psi$ behaves as an effective sound speed.

We shall  quantize the field $\Psi_{\bf k}$
by promoting it to the operator. We can use the mode expansion 
\begin{align}
 \hat{\Psi}_{\bf k}=
 v_{\bf k}(\tau)\hat{a}_{\bf k}+
 v^*_{\bf -k}(\tau)\hat{a}^\dagger_{\bf -k}  \ ,
\end{align}
where $\hat{a}_{\bf k}$ and $\hat{a}^\dagger_{\bf -k}$ are annihilation
and creation operators. The mode function  $v_{\bf k}$ satisfies the
same equation as $\Psi_{\bf k}$
\begin{eqnarray}
 v''_{\bf k}+\left(C_\psi^2k^2-\frac{A_\psi''}{A_\psi}\right) v_{\bf k}=0 \ .
 \label{eom_Psi}
\end{eqnarray}

Under the slow-roll conditions, $A_\psi$ can be approximated as
\begin{eqnarray}
 A_\psi^2 =
 6a^2 \left[
 1-\left(1-\rho^2/6-\sigma\right)
 \frac{1-\sigma/2}{1-3\sigma/2}
 \right]
 = a^2\rho^2
 = a^2(2\epsilon+4\alpha/3+\gamma/2)  \ ,
 \label{A}
\end{eqnarray}
where we have kept terms to the leading order in slow-roll parameters.
Similarly,  $C_\psi$ can be calculated as
\begin{eqnarray}
 C_\psi^2 =
 \frac{2}{A_\psi^2}\left[
 a\frac{{\rm d}}{{\rm d}t}
 \left(\frac{a}{H}\right)\frac{1-\sigma}{1-3\sigma/4}
 -a^2\left(1-\frac{\ddot{\xi}}{2M_{\rm Pl}^2}\right)
 \right]
 =
 \frac{2}{\rho^2}\left[
 \left(1-\frac{\dot{H}}{H^2}\right)(1-\sigma/4)
 -1+\frac{\ddot{\xi}}{2M_{\rm Pl}^2}
 \right] \ .
\end{eqnarray}
The last term can be negligible as is seen from
the following calculation
\begin{align*}
 \ddot{\xi}
 =\xi_{,\phi\phi}\dot{\phi}^2+\xi_{,\phi}\ddot{\phi}
 =M_{\rm Pl}^2
 H^2\xi_{,\phi\phi}\rho^2+H\dot{\xi}\frac{\ddot{\phi}}{H\dot{\phi}}
 =2M_{\rm Pl}^2\beta\rho^2+
 M_{\rm Pl}^2\sigma\frac{\ddot{\phi}}{H\dot{\phi}} \ .
\end{align*}
Thus, we have
\begin{align}
 C_\psi^2=\frac{2}{\rho^2}
 \left(
 -\frac{\dot{H}}{H^2}-\sigma/4
 \right)
 =\frac{2(\epsilon+\alpha/3+\alpha/3+\gamma/4)}
 {2\epsilon+4\alpha/3+\gamma/2}=1+{\cal O}(\epsilon) \ ,
\end{align}
where ${\cal O}(\epsilon)$ represents the first order corrections 
in slow-roll parameters.
Now, let us calculate the effective potential.  
Using the relation
\begin{align}
 \dot{\rho}
 =
 -M_{\rm Pl}\frac{V_{,\phi\phi}}{V}\dot{\phi}
 +M_{\rm Pl}\frac{V_{,\phi}^2}{V^2}\dot{\phi}
 -\frac{1}{6M_{\rm Pl}^3}V_{,\phi}\xi_{,\phi}\dot{\phi}
 -\frac{1}{6M_{\rm Pl}^3}V\xi_{,\phi\phi}\dot{\phi}
 =
 \left(
 2\epsilon-\eta-\frac{2}{3}\alpha-\beta
 \right)H\rho\ ,
\end{align}
we obtain
\begin{align}
 \frac{A'_\psi}{A_\psi}
 =\frac{a}{2}\frac{(A_\psi^2\dot{)}}{A_\psi^2}
 =a\left(H+\frac{\dot{\rho}}{\rho}\right)
 ={\cal H}
  \left(
 1+2\epsilon-\eta-\frac{2}{3}\alpha-\beta
 \right)\ .\nonumber
\end{align}
From Eq.(\ref{hdot2}), we deduce
\begin{align}
 \left(\frac{1}{\cal H}\right)'
 = a \frac{d}{dt}\left(\frac{1}{aH}\right)
 =- \frac{\dot{H}}{H^2}-1
 = -1-\epsilon-\frac{\alpha}{3} \ .
\end{align}
We integrate this equation to get  $\cal H$ as 
\begin{align}
 {\cal H}=\frac{1+\epsilon+\alpha/3}{-\tau}\ , 
 \label{conhub}
\end{align}
where we assumed $\tau <0 $. 
Hence, we reached the following result
\begin{align}
 \frac{A'_\psi}{A_\psi}=\frac{(1+\epsilon+\alpha/3)
 (1+2\epsilon-\eta-2\alpha/3-\beta)}{-\tau}
 =
 \frac{1+3\epsilon-\eta-\alpha/3-\beta}{-\tau}\ .
\end{align}
Thus, we can get the effective potential
\begin{align}
 \frac{A''_\psi}{A_\psi}
 =
 \frac{{\rm d}}{{\rm d}\tau}
 \left(\frac{A'_\psi}{A_\psi}\right)
 + \left(\frac{A'_\psi}{A_\psi}\right)^2
 =
 \frac{2+9\epsilon-3\eta-\alpha-3\beta}{\tau^2}
 =
 \frac{\nu_\psi^2-1/4}{\tau^2}\ ,
\end{align}
where we have defined
\begin{align}
 \nu_\psi\equiv
 \frac{3}{2}+3\epsilon-\eta-\alpha/3-\beta\ .
\end{align}
Now, Eq.(\ref{eom_Psi}) reduces to
\begin{align}
 v''_{\bf k}+\left(
 C^2_\psi k^2- \frac{\nu_\psi-1/4}{\tau^2}
 \right)v_{\bf k}=0 \ .
\end{align}
In the asymptotic past $\tau\rightarrow-\infty$, the effective potential vanishes.
 Hence, we can specify the Bunch-Davies vacuum by choosing
 the positive frequency mode function there as
\begin{align}
 v_{\bf k}=\frac{1}{\sqrt{2C_\psi k}}{\rm e}^{-iC_\psi k\tau}
 \quad
 \ .
\end{align}
It is easy to find the corresponding mode function  
\begin{align}
 v_{\bf k}(\tau)
 =\frac{\sqrt{-\pi\tau}}{2}
 {\rm e}^{i(\pi/4+\pi \nu_\psi/2)}
 {\rm H}^{(1)}_{\nu_\psi}(-C_\psi k\tau) \ ,
\end{align}
where ${\rm H}^{(1)}_{\nu_\psi}$ is the first kind of Hunkel function. 
Asymptotic form of $v_{\bf k}$  outside the horizon $-k\tau\ll1$
reads
\begin{align}
 v_{\bf k}(\tau)
 =\sqrt{\frac{-\tau}{2(-C_\psi k\tau)^3}}
 {\rm e}^{i(-\pi/4+\pi \nu_\psi/2)}
 \frac{\Gamma(\nu_\psi)}{\Gamma(3/2)}
 \left(\frac{-C_\psi k\tau}{2}\right)^{3/2-\nu_\psi}
 \ .
\end{align}
The power spectrum of scalar perturbation ${\cal P}_\psi$
can be defined as
\begin{align}
 \langle 0|\hat{\psi}^\dagger \hat{\psi}|0\rangle
 =
 \int{\rm d}({\rm log}k)
 \frac{1}{2\pi^2}k^3\frac{|v_{\bf k}|^2}{A_\psi^2}
 =
 \int{\rm d}({\rm log}k){\cal P}_\psi(k) \ .
\end{align}
Thus, we obtain 
\begin{align}
 {\cal P}_\psi(k)
 =\frac{1}{C_\psi^3}\frac{1}{4\pi^2}\frac{1}{A_\psi^2\tau^2}
 \left( \frac{\Gamma(\nu_\psi)}{\Gamma(3/2)}\right)^2
 \left(\frac{-C_\psi k\tau}{2}\right)^{3-2\nu_\psi} \ .
 \label{P_temp}
\end{align}
From Eqs.(\ref{A}) and (\ref{conhub}), we have
\begin{align}
 \frac{1}{A_\psi^2\tau^2}
 =
 \frac{{\cal H}^2}{a^2(2\epsilon+4\alpha/3+\gamma/2)(1+\epsilon+\alpha/3)^2}
 =
 \frac{H^2}{2\epsilon+4\alpha/3+\gamma/2}\ .
\end{align}
Using the above relation, we can simplify the power spectrum as
\begin{align}
 {\cal P}_\psi(k)
 =
 \frac{1}{2\epsilon+4\alpha/3+\gamma/2}
 \left(\frac{H}{2\pi}\frac{\Gamma(\nu_\psi)}{\Gamma(3/2)}\right)^2
 \left(\frac{-
 %C_\psi
 k\tau}{2}\right)^{3-2\nu_\psi}\ . 
\end{align}
Then, the scalar spectral index  $n_\psi$  is given by
\begin{align}
 n_\psi-1=3-2\nu_\psi=
 -6\epsilon+2\eta+\frac{2}{3}\alpha+2\beta \ .
 \label{n_psi}
\end{align}
Of course, in the absence of the Gauss-Bonnet term,  $\alpha=\beta=0$, 
we recover the conventional formula.

\section{Tensor Perturbations}

In this section, we compute the power spectrum for tensor perturbations.
In addition to the spectral index and the amplitude, we also calculate
the circular polarization of gravitational waves.

The metric we consider is
\begin{align}
 {\rm d}s^2
 =a^2(\tau)\left[-{\rm d}\tau^2
 +(\delta_{ij}+h_{ij}){\rm d}x^i{\rm d}x^j\right]  \ ,
\end{align}
where $h_{ij}$ is a transverse and traceless tensor.
Let us expand $h_{ij}$ in plain waves and circular polarization tensor
as
\begin{align}
 h_{ij}=\frac{\sqrt{2}}{M_{\rm Pl}}\sum_\pm
 \int\frac{{\rm d}k^3}{(2\pi)^3}\phi^\pm_{\bf k}
 {\rm e}^{i{\bf k\cdot x}}p^\pm_{{\bf k}, ij}\ ,
\end{align}
where $p^\pm_{{\bf k}, ij}$ denotes polarization tensor for
circular polarization, and $\pm$ denotes helicity of each mode. 
The action  for $\phi^\pm_{\bf k}$ becomes,
\begin{eqnarray}
 S=\sum_\pm\frac{1}{2}
 \int{\rm d}\tau\int\frac{{\rm d}k^3}{(2\pi)^3}
 A_{\rm  T}^2\left(
 |(\phi^\pm_{\bf k})'|^2
 -C^2_{\rm T} k^2|\phi^\pm_{\bf k}|^2
 \right) \ ,
\end{eqnarray}
where
\begin{align}
 A_{\rm T}^2&\equiv
 a^2\left(1-\sigma/2\mp\frac{1}{2M_{\rm Pl}^2}\frac{k\omega'}{a^2}\right)
 \label{Ah}
 ,\quad
 C_{\rm T}^2\equiv
 1+\frac{1}{2M_{\rm Pl}^2}
 \frac{{\cal H}\xi'}{A_{\rm T}^2}-\frac{1}{2M_{\rm Pl}^2}
 \frac{a^2\ddot{\xi}}{A_{\rm T}^2}
 \ .
\end{align}
Note that there exists helicity dependence both in $A_{\rm T}$ and $C_{\rm T}$,
although we use notation $A_{\rm T}$ and $C_{\rm T}$ rather than
$A_{\rm T}^\pm$ and $C_{\rm T}^\pm$, for simplicity.
Apparently, the parity violating term gives rise to the left-right asymmetry,
and hence the  circular polarization of primordial gravitational waves.
The observability of the circular polarization depends on the 
amplitude of the gravitational waves and the degree of the polarization.
Since the parity violating term does not affect
the background evolution and scalar perturbations,
the coupling function $\omega$  cannot be  constrained strongly.
Here, we only require the absence of the ghost gravitons.
From Eq.(\ref{Ah}), we see a serious problem would appear
 if the parity violating term become dominant. In that case, 
 $\pm$ sign appears as the overall factor of the action, and
 that means  the appearance of ghost in the gravitational waves.
 In other words, we cannot trust our treatment of the gravitational waves
 in this regime. 
 Hence, we impose that the 
term $|k\omega'/2a^2|$ must be lesser than $M_{\rm Pl}^2$ at 
$k/a=M_{\rm c}$, 
\begin{align}
1> \frac{M_{\rm c}}{2M_{\rm Pl}^2}\left|\frac{\omega'}{a}\right|
 =\frac{M_{\rm c}}{2M_{\rm Pl}}
 \left|\frac{\dot{\omega}}{M_{\rm Pl}}\right| \equiv \left|\Omega \right|
 \ ,
 \label{const1}
\end{align}
where $M_{\rm c}$ is the physical wave-number corresponding to the cut-off scale,
{\it e.g.} string scale, and we have defined 
\begin{align}
 \Omega\equiv\frac{1}{2}
 \frac{M_{\rm c}}{M_{\rm Pl}}\frac{\dot{\omega}}{M_{\rm Pl}}
 \simeq{\rm const.}    \ .
\end{align}
Here, since we are considering
slow-roll inflation, we assumed  
$\dot{\omega}=\omega'/a\simeq {\rm const.}$ in the period of interest.
The amplitude of gravitational waves is determined at the horizon-crossing, 
$k/a\simeq H$. 
Thus, the effect of the parity violating  term can be estimated as
\begin{align}
 \frac{1}{M_{\rm Pl}^2}\left|\frac{k\omega'}{2a^2}\right|
 \simeq
 \frac{H}{2M_{\rm Pl}^2}|\dot{\omega}|
 < \frac{H}{M_{\rm c}}\ ,
\end{align}
where we have used the constraint (\ref{const1}) in the last inequality.
In other words, we can treat the parity violating effect as small correction
 always, since the Hubble parameter is smaller than the cut-off scale.

Defining new variables $\mu^\pm_{\bf k}\equiv A_{\rm T}\phi^\pm_{\bf k}$,
we obtain the canonical action
\begin{eqnarray}
 S =\sum_\pm\frac{1}{2}
 \int{\rm d}\tau\int\frac{{\rm d}k^3}{(2\pi)^3}
 \left(
 |(\mu^\pm_{\bf k})'|^2
 -C^2_{\rm T} k^2|\mu^\pm_{\bf k}|^2
 +\frac{A''_{\rm T}}{A_{\rm T}}|\mu^\pm_{\bf k}|^2
 \right) \ .
\end{eqnarray}
Taking the variation of the action with respect to $\mu^\pm_{\bf k}$,
 we obtain the equation for the gravitational waves 
\begin{eqnarray}
 (\mu^\pm_{\bf k})''+
 \left(
 C_{\rm T}^2k^2
 -\frac{A_{\rm T}''}{A_{\rm T}}
 \right)
 \mu^\pm_{\bf k}=0 \ .
 \label{eom_Mu}
\end{eqnarray}

Now, we quantize gravitational waves 
\begin{align}
 \hat{h}_{ij}=\frac{\sqrt{2}}{M_{\rm Pl}}\sum_\pm
 \int\frac{{\rm d}k^3}{(2\pi)^3}
 \frac{\hat{\mu}^\pm_{\bf k}}{A_{\rm T}}
 {\rm e}^{i{\bf k\cdot x}}p^\pm_{{\bf k},ij}\ ,
\end{align}
by using the mode expansion
\begin{align}
 \hat{\mu}^\pm_{\bf k}=u^\pm_{\bf k}(\tau)\hat{a}^\pm_{\bf k}+
 u^{\pm*}_{\bf -k}(\tau)\hat{a}^{\pm\dagger}_{\bf -k}\ .
\end{align}
Let us represent $A_{\rm T}$ and $C_{\rm T}$ in terms of slow-roll 
parameters as
\begin{eqnarray}
 A_{\rm T}^2
 =a^2\left(1+\frac{2}{3}\alpha+\frac{\gamma}{2}
 \mp\frac{k\Omega}{M_{\rm c}a}
 \right)
 \label{A_without}
\end{eqnarray}
and
\begin{eqnarray}
 C_{\rm T}^2=1+\frac{1}{2M_{\rm Pl}^2}\frac{{\cal H}\xi'}{A_{\rm T}^2}
 =1+\frac{\sigma}{2(1-\sigma/2\mp k\Omega/M_{\rm c}a)}
 =1-\frac{2}{3}\alpha-\frac{1}{2}\gamma\ .
\end{eqnarray}
Remember that $|\Omega|<1$  from the constraint (\ref{const1}).
Now, using the following relation
\begin{align}
 \frac{A'_{\rm T}}{A_{\rm T}}=\frac{1}{2}\frac{(A_{\rm T}^2)'}{A_{\rm T}^2}
 ={\cal H}\pm\frac{k\Omega}{2M_{\rm c}}\frac{\cal H}{a}
 =\frac{1+\epsilon+\alpha/3}{-\tau}
 \pm k\frac{H}{2M_{\rm c}}\Omega\ ,
\end{align}
we obtain the effective potential
\begin{align}
 \frac{A''_{\rm T}}{A_{\rm T}}
 =
 \frac{{\rm d}}{{\rm d}\tau}
 \left(\frac{A'_{\rm T}}{A_{\rm T}}\right)
 + \left(\frac{A'_{\rm T}}{A_{\rm T}}\right)^2
 =\frac{2+3\epsilon+\alpha}{\tau^2}
 \pm\frac{k}{-\tau}\frac{H}{M_{\rm c}}\Omega
 =\frac{\nu_{\rm T}^2-1/4}{\tau^2}
 \pm\frac{k}{-\tau}\frac{H}{M_{\rm c}}\Omega \ ,
\end{align}
where we have defined 
\begin{eqnarray}
   \nu_{\rm T}\equiv\frac{3}{2}+\epsilon+\frac{\alpha}{3} \ .
\end{eqnarray}
Thus, the mode function $u_{\bf k}$ obeys
\begin{align}
(u^\pm_{\bf k})''+
 \left(
 C_{\rm T}^2k^2
 -\frac{\nu_{\rm T}^2-1/4}{\tau^2} \mp\frac{k}{-\tau}
 \frac{H}{M_{\rm c}}\Omega
 \right)
 u^\pm_{\bf k}=0\ . \label{tensor_eom}
\end{align}
The Bunch-Davis vacuum is defined by the asymptotic form
of the mode function at $\tau\rightarrow-\infty $
\begin{align}
 u^\pm_{\bf k}
 =\frac{1}{\sqrt{2C_{\rm T} k}}{\rm e}^{-iC_{\rm T} k\tau}
    \ .
\end{align}
Using the confluent hypergeometric function ${\rm U}$, 
the corresponding positive frequency mode function $u^\pm_{\bf k}$ is analytically
given by
\begin{align}
 u^\pm_{\bf k}(\tau)
 &=
 {\rm e}^{-iC_{\rm T}k\tau}(-2C_{\rm T}k\tau)^{\nu_{\rm T}}
 \sqrt{-\tau}
 {\rm e}^{-i(\pi/4+\pi\nu_{\rm T}/2)}
 {\rm U}\left(
 \frac{1}{2}+\nu_{\rm T}\mp i\frac{1}{2C_{\rm T}}
 \frac{H}{M_{\rm c}}\Omega
 ,\ 
 1+2\nu_{\rm T}
 ,\ 
 2iC_{\rm T}k\tau
 \right)
 \nonumber \\
 &\qquad\qquad\qquad
 \times
 \exp\left(
 \pm\frac{\pi}{4C_{\rm T}}\frac{H}{M_{\rm c}}\Omega
 \right) 
 \ .
\end{align}
The asymptotic form of $u^\pm_{\bf k}$  outside the horizon $-C_{\rm T}k\tau\ll1 $
becomes
\begin{align}
 u^\pm_{\bf k}(\tau)
 =\sqrt{\frac{-\tau}{2(-C_{\rm T}k\tau)^3}}
 {\rm e}^{i(-\pi/4+\pi \nu_{\rm T}/2)}
 \frac{\Gamma(\nu_{\rm T})}{\Gamma(3/2)}
 \left(\frac{-C_{\rm T}k\tau}{2}\right)^{3/2-\nu_{\rm T}}
 \exp\left(
 \pm\frac{\pi}{4C_{\rm T}}\frac{H}{M_{\rm c}}\Omega
 \right)
 \ .  
\end{align}
Now, let us define the power spectrum ${\cal P}_{\rm T}$ as 
\begin{align}
 \langle 0|\hat{h}_{ij}\hat{h}^{ij}|0\rangle=
 \sum_\pm
 \int{\rm d}(\log k)
 \frac{2}{\pi^2}k^3\frac{|u^\pm_{\rm k}|^2}{A_{\rm T}^2}
 =
 \sum_\pm
 \int{\rm d}(\log k)
 {\cal P}^\pm_{\rm T}
 =
 \int{\rm d}(\log k)
 {\cal P}_{\rm T}
 \ .
\end{align}
Thus, we have the formula ${\cal P}^\pm_{\rm T}$ for each polarization
\begin{align}
 {\cal P}^\pm_{\rm T}
 &=\frac{1}{C_{\rm T}^3}\frac{1}{\pi^2}\frac{1}{A_{\rm T}^2\tau^2}
 \left( \frac{\Gamma(\nu_{\rm T})}{\Gamma(3/2)}\right)^2
 \left(\frac{-C_{\rm T}k\tau}{2}\right)^{3-2\nu_{\rm T}}
 \exp\left(
 \pm\frac{\pi}{2C_{\rm T}}\frac{H}{M_{\rm c}}\Omega
 \right)
 \nonumber \\
 &=
 \frac{4}{1+\epsilon-\alpha-\gamma}
 \left(\frac{H}{2\pi} \frac{\Gamma(\nu_{\rm T})}{\Gamma(3/2)}\right)^2
 \left(\frac{-C_{\rm T}k\tau}{2}\right)^{3-2\nu_{\rm T}}
 \left(
 1\pm \frac{\pi}{2}\frac{H}{M_{\rm c}}\Omega
 \right)
\end{align}
The total power spectrum ${\cal P}_{\rm T}={\cal P}^+_{\rm T}+{\cal P}^-_{\rm T}$
becomes
\begin{eqnarray}
 {\cal P}_{\rm T}  =
 \frac{8}{1+\epsilon-\alpha-\gamma}
 \left(\frac{H}{2\pi} \frac{\Gamma(\nu_{\rm T})}{\Gamma(3/2)}\right)^2
 \left(\frac{-C_{\rm T}k\tau}{2}\right)^{3-2\nu_{\rm T}}
 \ .
\end{eqnarray}
It is easy to read off the tensor spectral index $n_{\rm T}$ as
\begin{align}
 n_{\rm T}=3-2\nu_{\rm T}
 =-2\epsilon-\frac{2}{3}\alpha\ .
 \label{nT}
\end{align}
Again, this result reduces to the conventional slow-roll formula when
$\alpha=0$.
The most impressive feature of the Gauss-Bonnet effects in the
slow-roll inflation is shown in Eq.(\ref{nT}). 
In the usual case, because  $\epsilon>0$,
$n_{\rm T}$ takes a negative value, namely, the tensor spectrum is red.
However, if we incorporate the effect of the Gauss-Bonnet term, tensor
spectrum could be blue, because $\alpha$ takes either positive or 
negative value.  Hence, detection of the blue spectrum in
the gravitational waves through the observation of B-mode polarization
might indicate the existence of the Gauss-Bonnet term. 
In the case of the blue spectrum,  $n_{\rm T}>0$, 
from Eq.(\ref{hdot}), we see that the background evolution
 corresponds to superinflation $\dot{H}>0$. 
In this case, inflaton field could climb up the potential slope, rather
than roll down. This interesting possibility occurs due to
the Gauss-Bonnet induced effective potential.

Since we have obtained both the scalar and tensor spectrum, 
we can deduce the tensor-to-scalar ratio $r$:
\begin{align}
 r\equiv\frac{{\cal P}_{\rm T}}{{\cal P}_\psi}
 &= 8\frac{2\epsilon+4\alpha/3+\gamma/2}{1+\epsilon-\alpha-\gamma}
 \left(\frac{-k\tau}{2}\right)^{2\nu_\psi-2\nu_{\rm T}}
 =
 \left(
 16\epsilon+\frac{32}{3}\alpha+4\gamma
 \right)
 \left(\frac{k/a}{2H}\right)^{n_{\rm T}-(n_\psi-1)}
 \ .
\end{align}
Since we have the approximate relation $n_\psi-1=n_{\rm T}$, 
the last factor must be 1.
 Hence, the tensor-to-scalar ratio $r$ is given by
\begin{align}
 r
 \sim
 16\epsilon+\frac{32}{3}\alpha+4\gamma
 =
 16\epsilon+\frac{32}{3}\alpha+\frac{16}{9}\frac{\alpha^2}{\epsilon}
 \geq \frac{32}{3}\left( \alpha + |\alpha | \right)
 \ . 
\end{align}  
For a negative $\alpha$, we have the minimum, that is,  $r=0$.
In the case $1 \geq \alpha \geq \epsilon$, the last term dominates
and give rise to the sizable tensor-to-scalar ratio
even if $\epsilon$ is extremely small.

Another interesting observable is the circular polarization $\Pi$,
which is defined as difference between left and right helicity modes
\begin{align}
 \Pi\equiv\frac{{\cal P}^+_{\rm T}-{\cal P}^-_{\rm T}}
 {{\cal P}^+_{\rm T}+{\cal P}^-_{\rm T}}
 =\frac{\pi}{2}\frac{H}{M_{\rm c}}\Omega\ .
\end{align}
One can see that the  circular polarization  is 
small under the constraint  (\ref{const1}).
However, it is important to quantify how small it is. Let us consider 
some extreme case, $H=10^{-4}M_{\rm Pl}$, $M_{\rm c}=10^{-3}M_{\rm Pl}$,
$\Omega=0.1$, then, the ratio becomes $\Pi\simeq 0.015$.
This means we have a hope to measure the sub-percent order of polarization due to
the parity violating term~\cite{Saito:2007kt,Seto:2006hf,taruya}. 
It should be stressed that
the effect of the Gauss-Bonnet term could enhance the amplitude of
the primordial gravitational waves. That also enhances the detectability.

\section{Observational implications}

In the previous sections, we have derived general formula for observables
in slow-roll inflation with higher curvature corrections. 
Here, we will discuss several models as an illustration. 

Now, let us consider chaotic inflation models with higher 
curvature corrections.  We take the functions as
\begin{align}
 V=\frac{1}{2}m^2\phi^2\ , \quad
 \xi=\lambda{\rm e}^{-\kappa\phi/M_{\rm Pl}} \ ,
 \label{example}
\end{align}
where $m$ is the mass of the inflaton and $\lambda , \kappa$ are parameters
of the coupling function $\xi$.
Then, the slow-roll parameters can be calculated as
\begin{align}
 \epsilon=\eta=2\frac{M_{\rm Pl}^2}{\phi^2},\quad
 \alpha=-\frac{\kappa\lambda}{4}\frac{m^2\phi}{M_{\rm Pl}^3}
 {\rm e}^{-\kappa\phi/M_{\rm Pl}},
 \quad
 \beta=\frac{\kappa^2\lambda}{12}\frac{m^2\phi^2}{M_{\rm Pl}^4}
 {\rm e}^{-\kappa\phi/M_{\rm Pl}}
 ,\quad
 \gamma=\frac{\kappa^2\lambda^2}{72}
 \frac{m^4\phi^4}{M_{\rm Pl}^8}{\rm e}^{-2\kappa\phi/M_{\rm Pl}}
 \ .
\end{align}
\begin{figure}[htbp]
 \centering
 \includegraphics[scale=0.6]{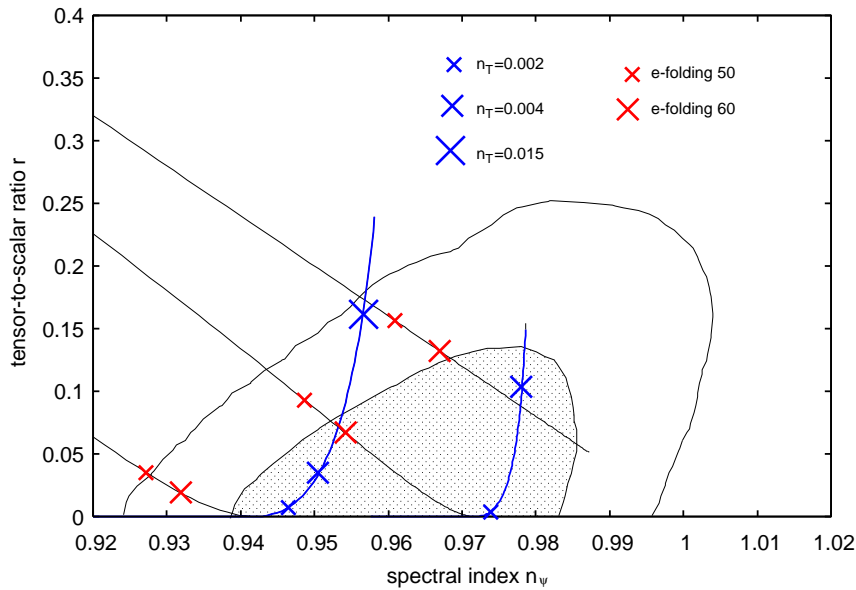}
 \caption{Expected spectral index $n_\psi$ and tensor-to-scalar ratio
 $r$ of the model (\ref{example}) are shown. We have also plotted 
 constraints  from WMAP 5-year result, combined with BAO and Type I SN.
 The contours denote 68\% and 95\% confidence level. We used 
 the out-put from Cosmological Parameters Plotter at LAMBDA~\cite{lambda}.
 Each line corresponds
 to the parameter (i) $m=10^{-5}\times M_{\rm Pl}$, $\lambda=0$, $\kappa=0$ 
 (ordinary slow-roll), (ii) $m=10^{-5}\times M_{\rm Pl}$, $\lambda=10^{10}$,
 $\kappa=0.003$ and (iii) $m=10^{-5}\times M_{\rm Pl}$, $\lambda=10^{10}$, $\kappa=0.01$,
 from upper to lower, taking $\phi>0$.
 Blue-colored lines denote the region that generate blue spectrum in
 tensor modes, and we plot some values of $n_{\rm T}$. }
 \label{ns_r}
\end{figure}
We can evaluate the spectral index and the tensor-to-scalar ratio.
 The results are shown in FIG.\ref{ns_r}.
 There, we have plotted the scalar spectral index $n_\psi$ and tensor-to-scalar ratio
 $r$ of the model (\ref{example}). We have also plotted 
 constraints  from WMAP 5-year result, combined 
 with baryon acoustic oscillations (BAO) and Type I supernova (SN).
 It turns out that our simple example explains WMAP results 
better than the conventional slow-roll inflation scenario.
 We see these observables are sensitive to higher curvature corrections.
 Hence, it implies that the higher curvature corrections are 
 relevant to precision cosmology.
 Here, we notice that there exist parameter regions represented by blue lines
  which leads to the blue spectrum in tensor modes.
  We plotted some typical values of $n_{\rm T}$ in Fig.\ref{ns_r}.
   The e-folding numbers for the blue region  cannot be estimated,
 because they correspond to superinflation. In that case, the model (\ref{example})
  must be modified so that the inflation ends.

\begin{figure}[htbp]
 \centering
 \includegraphics[scale=0.6]{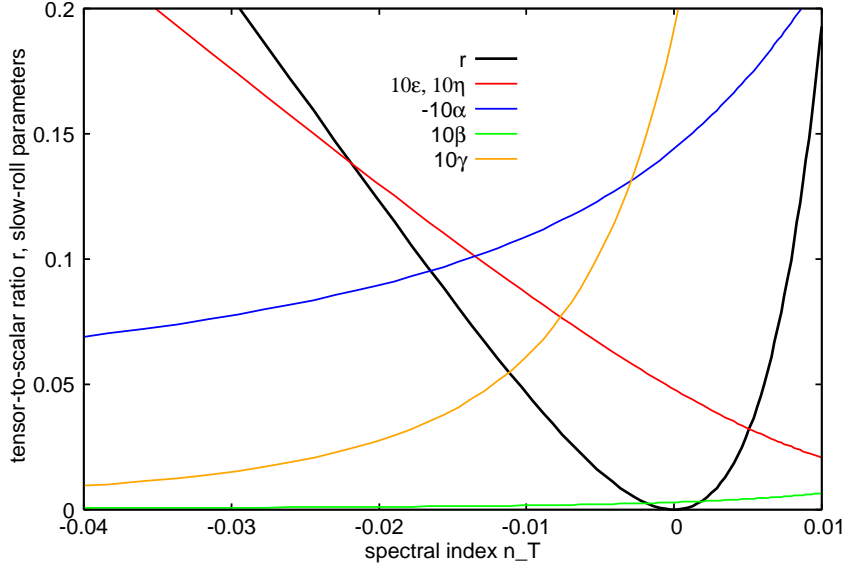}
 \caption{The tensor to the scalar ratio and the tensor spectral index 
  are plotted. We have also shown the slow-roll parameters.
 Here we set parameters as $m=10^{-5}\times M_{\rm Pl}$, $\lambda=10^{10}$ and
 $\kappa=0.003$.}
 \label{nt_r}
\end{figure}

With the higher curvature corrections, there is no consistency
relation as in the conventional inflation models. 
Therefore,
in Fig.\ref{nt_r}, we have plotted the $r-n_{T}$ diagram.
Since we have considered the negative $\alpha$, the minimum of the
tensor-to-scalar ratio is zero. Taking look at values for
slow-roll parameters, we see the Gauss-Bonnet term dominates 
in the region $n_T \geq -0.01$.

\begin{figure}[htbp]
 \centering
 \includegraphics[scale=0.6]{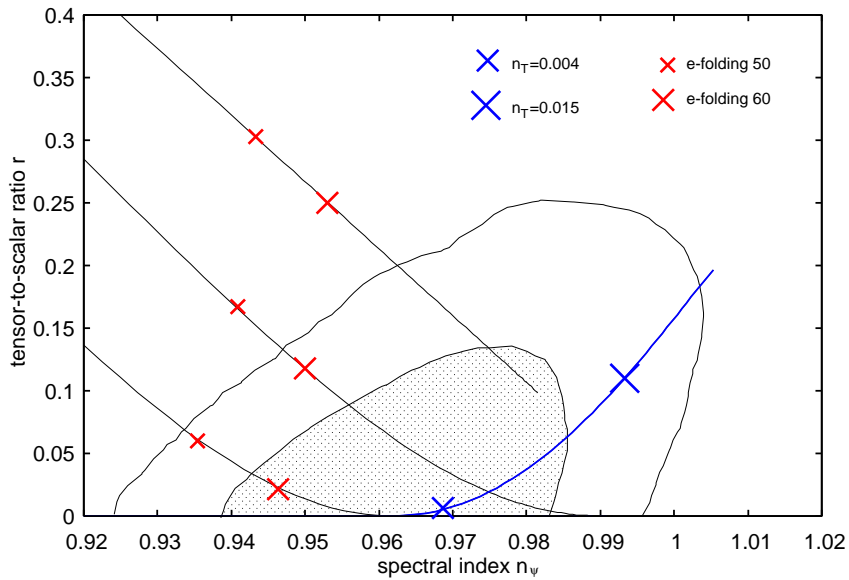}
 \caption{Expected spectral index $n_\psi$ and tensor-to-scalar ratio
 $r$ of the model (\ref{example_2}) are plotted.
 The constraints from WMAP 5-year data are also shown. 
 Each line corresponds
 to the parameter (i) $\Lambda=10^{-14}$, $\lambda=0$, $\kappa=0$ 
 (ordinary slow-roll), (ii) $\Lambda=10^{-14}$, $\lambda=1.5\times10^{10}$,
 $\kappa=0.1$ and (iii) $\Lambda=10^{-14}$, $\lambda=3.5\times10^{10}$, 
 $\kappa=0.1$,
 from upper to lower.}
 \label{ns_r_phi4}
\end{figure}

We consider $\phi^4$ model as another example. We 
parameterize the potential and the coupling functions as
\begin{align}
 V=\Lambda\phi^4,
 \quad
 \xi=\lambda{\rm e}^{-\kappa \phi/M_{\rm Pl}} \ , 
 \label{example_2}
\end{align}
where $\Lambda$ is a coupling constant.
Then, the slow-roll parameters read
\begin{align}
 \epsilon=8\frac{M_{\rm Pl}^2}{\phi^2},\quad
 \eta=12\frac{M_{\rm Pl}^2}{\phi^2},\quad
 \alpha=-\kappa\lambda\Lambda \frac{\phi^3}{M_{\rm Pl}^3}
 {\rm e}^{-\kappa\phi/M_{\rm Pl}},
 \quad
 \beta=\frac{\kappa^2\lambda\Lambda}{6}\frac{\phi^4}{M_{\rm Pl}^4}
 {\rm e}^{-\kappa\phi/M_{\rm Pl}}
 ,\quad
 \gamma=\frac{\kappa^2\lambda^2\Lambda^2}{18}
 \frac{\phi^8}{M_{\rm Pl}^8}{\rm e}^{-2\kappa\phi/M_{\rm Pl}}
 \ .
\end{align}
The scalar spectral index and the tensor-to-scalar ratio
for some parameters are shown in FIG.\ref{ns_r_phi4}.
We see that the $\phi^4$ model with no higher curvature corrections cannot 
explain observations at all. However, if we take into account the higher curvature
effects, the result changes  and the $\phi^4$ model seems to survive.

\begin{figure}[htbp]
 \centering
 \includegraphics[scale=0.6]{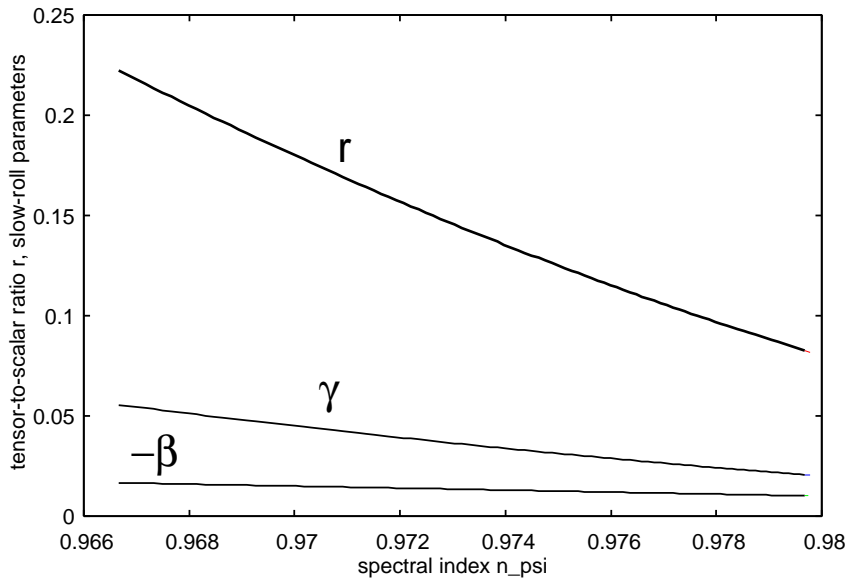}
 \caption{The tensor-to-scalar ratio $r$ and slow-roll parameters 
 of the model (\ref{v_flat}) corresponding to
 scalar spectral index $n_\psi$  are plotted. 
 We set parameters as $V_0=10^{-10}\times M_{\rm Pl}^4$, 
 $\lambda=-10^{11}$ and $\kappa=0.1$.}
 \label{enhance}
\end{figure}

Let us give the last example where the tensor-to-scalar ratio
$r$ is enhanced due to the Gauss-Bonnet term.
To highlight the feature, let us consider the flat potential model
\begin{align}
 V=V_0,\quad
 \xi=\lambda{\rm e}^{-\kappa\phi/M_{\rm Pl}}\ ,
 \label{v_flat}
\end{align}
where $V_0$ is a constant. 
For this model, the slow-roll parameters are given by
\begin{align}
 \epsilon=\eta=\alpha=0,\quad
 \beta=\frac{\kappa^2\lambda}{6}
 \frac{V_0}{M_{\rm Pl}^4}{\rm e}^{-\kappa\phi/M_{\rm Pl}},\quad
 \gamma=\frac{\kappa^2\lambda^2}{18}\frac{V_0^2}{M_{\rm Pl}^8}
 {\rm e}^{-2\kappa\phi/M_{\rm Pl}}\ .
\end{align}
It is clear that $r$ becomes zero and scalar spectrum becomes completely
flat, if the higher curvature corrections are absent.
Apparently, the spectral index $n_\psi =1$ is not the favored value.
Moreover, the detection of the primordial gravitational waves would 
observationally falsify this model. 
However, the higher curvature terms change the situation, since $\beta$
and $\gamma$ are not zero!
We show $r - n_\psi$ diagram, with correction terms, in FIG.\ref{enhance}.
Due to the higher curvature terms, 
the tensor-to-scalar ratio $r$ becomes detectable and 
the scalar spectral index $n_\psi$ is now a preferable value. 
This result clearly indicates the importance of higher curvature effects.

\section{Conclusion}

We have studied the slow-roll inflationary scenario with
the leading corrections, namely, the Gauss-Bonnet and the parity
violating terms. We have obtained the 
 higher curvature corrections to the scalar spectral index,
 the tensor spectral index, and the tensor-to-scalar ratio. 
We found that the higher curvature corrections can not be negligible in the dynamics
of the scalar field, although they are negligible in the Friedmann
equation. 
It turned out that the tensor-to-scalar ratio could be enhanced
because of the modified dynamics of the scalar field. 
We have  found the tensor spectral index could be blue due to the
 Gauss-Bonnet term.
We have also calculated the degree of circular polarization  of gravitational
waves generated during the  slow-roll inflation. 
We discussed that the circular polarization can be observable 
 due to the Gauss-Bonnet and parity violating terms.
 We also presented several examples to show
  observational implications of higher curvature corrections 
 for chaotic inflationary models. 
 We revealed that the observables  are sensitive to the higher curvature
 corrections. Interestingly, there are some models which shows the slowly
 climbing up of the potential. Although it is an exotic possibility,
 the phenomena arises within the slow-roll approximation. 
The climbing up inflationary model deserves further investigation.

 We can extend the present analysis in various ways. 
 Although we have considered only the scalar field, 
 we could have incorporated  the vector field into 
 the system~\cite{Dimopoulos:2008rf,Yokoyama:2008xw,Kanno:2008gn},
 which leads to more general higher derivative corrections.
 However, the most important issue is to determine the coupling functions
 based on the fundamental physics such as the superstring theory.
 We leave these issues for future work.

\begin{acknowledgements}
We would like to thank Sugumi Kanno for the collaboration in the early
 stage of this project.
J.S. is supported by the Japan-U.K. Research Cooperative Program, 
  Grant-in-Aid for  Scientific
Research Fund of the Ministry of Education, Science and Culture of Japan 
 No.18540262 and No.17340075.  
\end{acknowledgements}

\end{document}